\documentclass[fleqn,10pt]{wlscirep}
\usepackage[utf8]{inputenc}
\usepackage[T1]{fontenc}
\usepackage{graphicx}
\usepackage{dcolumn}
\usepackage{bm}
\usepackage{epstopdf}
\usepackage{amssymb}

\title{Demonstration of a MOT in a Sub-Millimeter Membrane Hole}

\author[1,*]{Jongmin Lee}
\author[1]{Grant Biedermann}
\author[1]{John Mudrick}
\author[1]{Erica A. Douglas}
\author[1]{Yuan-Yu Jau}

\affil[1]{Sandia National Laboratories, Albuquerque, New Mexico 87185, USA}
\affil[*]{jongmin.lee@sandia.gov}

\begin{abstract}
We demonstrate the generation of a cold-atom ensemble within a sub-millimeter diameter hole in a transparent membrane, a so-called ``membrane MOT''. With a sub-Doppler cooling process, the atoms trapped by the membrane MOT are cooled down to 10\,$\mu$K. The atom number inside the unbridged/bridged membrane hole is about $10^4$ to $10^5$, and the $1/e^2$-diameter of the MOT cloud is about 180\,$\mu$m for a 400\,$\mu$m-diameter membrane hole. Such a membrane device can, in principle, efficiently load cold atoms into the evanescent-field optical trap generated by the suspended membrane waveguide for strong atom-light interaction and provide the capability of sufficient heat dissipation at the waveguide. This represents a key step toward the photonic atom trap integrated platform (ATIP)\,\cite{Gehl21}.
\end{abstract}

\begin{document}

\flushbottom
\maketitle
%
%
\thispagestyle{empty}


\section*{Introduction}

The past two decades have witnessed remarkable advances in computing and sensing with demonstrations that leverage coherence and entanglement in systems well-described by quantum mechanics \,\cite{Weiss17, Alzar19}.  Concurrently, the advent of microfabrication techniques promises new frontiers in quantum applications\,\cite{DiVincenzo00, Zoller01, Qi18} through atom-light interactions\,\cite{Mabuchi06, Stern13, Rolston13a, Lukin14, Kimble15, Rolston15, Stievater16, Black18, Pfau18, Hung19, Hackermueller20, Kimble20} such as incorporating compact magneto-optical traps (MOTs)\,\cite{Takuma91, Jhe96, Hinds09, Arnold10, Hinds13, Lee13, Bongs16, Squires17, Arnold17} on atom chips\,\cite{Hansch99, Schmiedmayer00, Folman05, Anderson05, Reichel07, Anderson13,Himsworth14, Folman16} and superconducting circuits\,\cite{Schmiedmayer09, Rolston11, Rolston13b, Schmiedmayer15, Park15, Fortagh17}.  Quantum engineering of integrated quantum systems has seen the development of compact and scalable laser systems using hybrid integrated photonic circuits\,\cite{Kodigala19, Zwiller20} with silicon photonics, III-V photonics, and nonlinear optics. Such techniques are poised to facilitate reliable operation of cold-atom positioning, navigation and timing (PNT) sensors\,\cite{Biedermann14} in new and challenging environments. Engineering of quantum systems has also enabled integrated, on-chip quantum computing platforms capable of individual spin addressing, spin-spin entanglement and spin readout. Quantum processors with more than 50 qubits have been demonstrated with superconducting circuits\,\cite{Knight17, Martinis19}.  Likewise, new efforts are exploring novel surface ion-trap platforms \,\cite{Zhang17, Chiaverini20} combining microfabricated surface electrodes and integrated photonics on the same chip. These quantum engineering efforts aim to implement a quantum-to-classical interface\,\cite{Reilly15} and leverage conventional computing and control systems to achieve optimum performance of integrated quantum systems across multiple platforms.

\begin{figure}[b!]
\includegraphics[width=1\columnwidth]{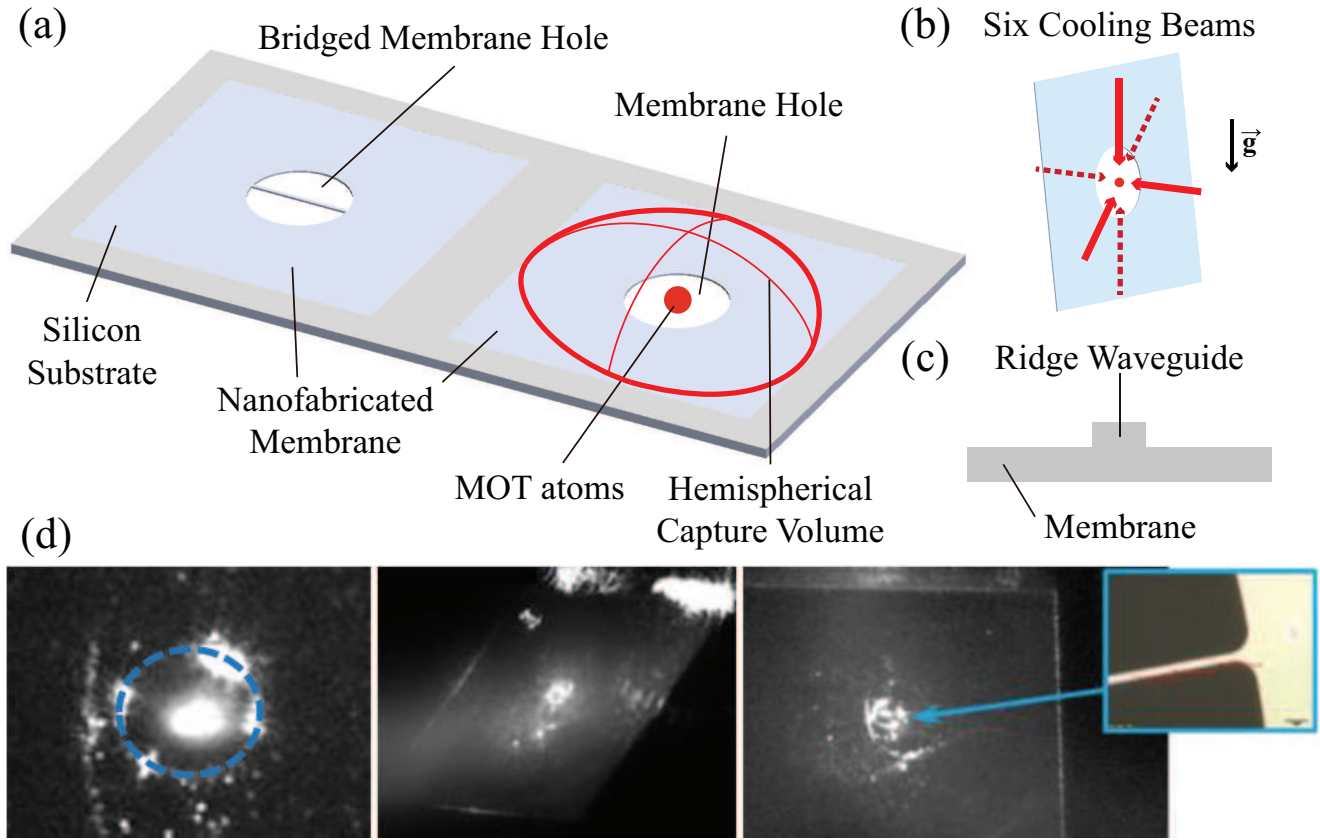}
\caption{(a) Concept of membrane MOT devices. The area of the membrane is 5 mm $\times$ 5 mm, and cold atoms in a membrane MOT can be generated in a sub-millimeter diameter membrane hole using six 3.8\,mm diameter cooling beams. The bridged membrane hole has a dummy waveguide which imitates a 3-$\mu\rm{m}$-width membrane ridge waveguide. (b) Six cooling beam configuration for a membrane MOT. (c) Concept of the membrane ridge waveguide. (d) (Left) Image of laser-cooled atoms at the center of a membrane hole without a dummy waveguide. (Middle, Right) Images of laser-cooled atoms at a bridged membrane hole. The diameter of the unbridged/bridged membrane hole is 400\,$\mu$m in both.}
\label{fig_1}
\end{figure}

Membrane MOT devices compatible with photonic integrated circuits (PIC) technology\,\cite{Ayi-Yovo20} can pave the way toward atom trap integrated platforms (ATIP)\,\cite{Gehl21} for neutral atoms. In this architecture, the suspended membrane waveguide can be used to trap and probe neutral atoms through the evanescent fields of optical waveguide modes, which is capable of delivering sufficient optical powers in vacuum while the membrane structure attached to the silicon substrate dissipates the heat generated from optical absorption in the waveguide. This heat dissipation capability could eliminate the need for the fabrication of optical waveguides on the substrate\,\cite{Rolston15, Stievater16, Black18}.  Photonic ATIPs will be crucial for enabling future neutral atom quantum applications. Atomic spins positioned near the waveguide surface offer an interface between spin information and the guided optical mode which can be processed in the PIC.  Compared to artificial counterparts\,\cite{Englund16,Popkin16},  neutral atoms offer powerful and compelling advantages in terms of homogeneous physical properties and long coherence and life times due to being well-isolated from noise sources. Compared to trapped ion approaches, neutral atoms offer near-term scaling advantages, larger atomic ensembles and a wide range of sensing modalities.
 
In this paper, we demonstrate a foundational technique for laser cooling atoms directly in the vicinity of an optical waveguide that can act as an information bus.  This approach is advantageous in terms of technical simplicity  and efficacy.  In particular, efficient atom loading near the suspended membrane waveguide is crucial for positioning and coupling many atoms to the evanescent-field guided mode.  In doing so, we can bring the success of evanescent-field optical traps around a nanofiber\,\cite{RB10, Lee15, Polzik17, Aoki19} to the photonic ATIP with unique features and new possibilities. Underpinning our approach is our introduction of a MOT produced in a sub-millimeter diameter hole on a microfabricated transparent membrane.  The membrane itself can support an optical waveguide that traverses the hole, allowing the optical field to interact with the atomic spins via the evanescent field extending into the vacuum.  In our demonstration, we imitate such a waveguide with a  three-micrometer-width mechanical beam fabricated from the membrane material, allowing us to test the atom loading dynamics around such a structure.
 
In our system\,\cite{Jau17}, we reimagine the interface between a MOT and a device with a transparent membrane that divides the MOT loading volume in two and collects cold atoms in a sub-millimeter hole in the membrane. According to our simulations (see Fig. 20 in Ref.\,\cite{Jau17}), through laser cooling, atoms dissipate kinetic energy and relax into the center of the trap without entering the other hemisphere.  Thus, many atoms eventually accumulate at the center of the membrane hole. The size or localization of membrane MOT atoms is limited by the size of the membrane hole, which can be readily sub-millimeter, whereas the size of the membrane itself can be orders of magnitude larger and define a generous loading volume. Using the membrane MOT with the 400\,$\mu$m-hole diameter, we loaded about $10^5$ atoms into the hole. Furthermore, this geometry is compatible with sub-Doppler cooling mechanisms\,\cite{Dalibard89, Chu89, Metcalf90, Phillips92} and we observe $\sim$10\,$\mu$K temperature in our setup.  Our membrane MOT approach offers a convenient means for positioning a large number of trapped atoms around microfabricated light guiding structures in the plane of an ATIP.

\section*{Experimental setup and membrane fabrication}

\begin{figure}[b!]
\includegraphics[width=1\columnwidth]{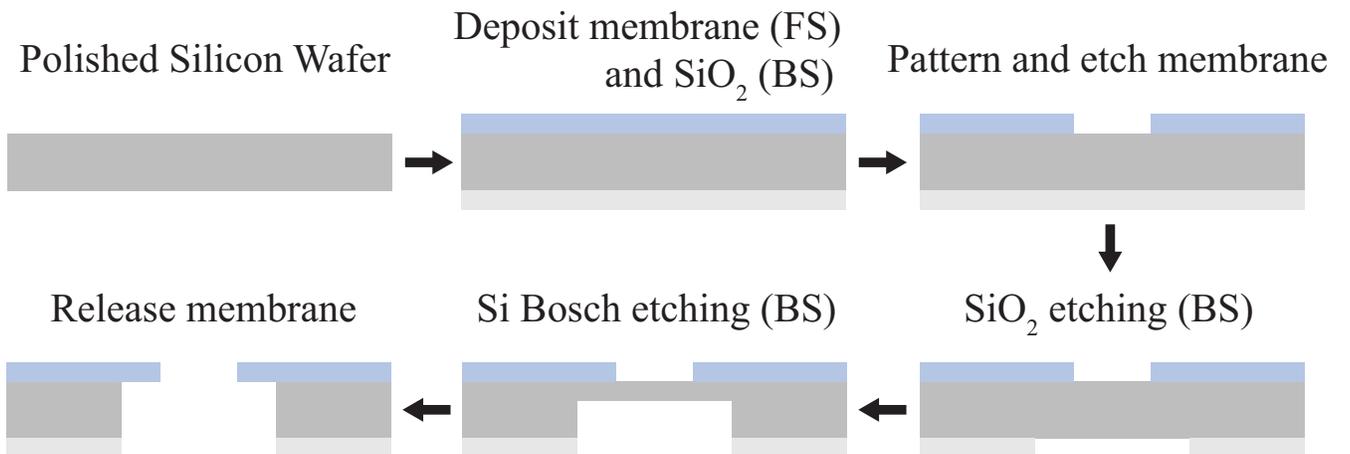}
\caption{Membrane fabrication process of SiN and AlN; FS is the front side, the top of a silicon wafer, and BS is the back side, the  bottom of a silicon wafer. The thickness of membranes is 200\,nm, and the membrane transmittance is greater than $95\%$ at 852nm wavelengths (laser cooling light for Cs atoms).}
\label{fig_fabrication}
\label{fig_2}
\end{figure}

In the experiment, $^{133}$Cs atoms were used for the free-space and membrane MOTs with cooling and repump beams (852\,nm, D2 transition) and an absorption probe (895\,nm, D1 transition). The vacuum setup was configured with a glass chamber (40\,mm$\times$40\,mm$\times$100\,mm), a mini-cube and a 5\,L/s ion pump, resulting in a pressure of $10^{-8}$\,mbar. The entire chamber, and hence the samples, were moved with a 1D lab-jack and a 2D translation stage and aligned to the free-space MOT. Six intensity-balanced collimated MOT beams (< $4\,\rm{mm}$ $1/e^2$ diameter) pass through the transparent membrane (5 mm $\times$ 5mm) and can be precisely aligned to the membrane hole, i.e., the MOT loading zone, using multiple 2D translation stages. As shown in Fig.\,\ref{fig_1}(b), four horizontal laser cooling beams cool atoms in the horizontal plane, and two vertical laser cooling beams cool atoms along the vertical axis, i.e., the gravity direction. The transparent membrane captures atoms with two large hemispherical capture volumes as shown in Fig.\,\ref{fig_1}(a) and loads MOT atoms into the sub-millimeter membrane hole. Two rods support the aluminum membrane sample holder with $\sim$ 40 degree angle, allowing the MOT beams to pass through the membrane without being occluded by the silicon substrate. The symmetry axis of quadrupole MOT coils is aligned to the vertical axis, i.e., the gravity direction. We checked Doppler-cooling and sub-Doppler cooling of the free-space MOT and the membrane MOT with and without a dummy waveguide at the center hole.

We used two membranes made of AlN (aluminum nitride) and SiN (silicon nitride) for the experiments. Importantly, the thermal conductivity of AlN is 10 times higher than SiN, which can be advantageous in terms of heat dissipation of a suspended waveguide. The absorption loss of membrane ridge waveguides (Fig.\,\ref{fig_1}(c)) needs to be considered because it is the main cause of heat generation at the suspended waveguide section. The fabrication process of the membrane MOT devices \,\cite{Jau17} is shown in Fig.\,\ref{fig_2}. Tensile SiN/AlN films were deposited on the front side of a silicon wafer. Front-side SiN was deposited via low pressure chemical vapor deposition (LPCVD) while AlN was deposited via reactive sputter physical vapor deposition (PVD). $\rm{SiO_{2}}$(silicon dioxide) was deposited via plasma enhanced chemical vapor deposited (PECVD) on the back side of the wafer to be used as a hard mask for Si Bosch etching. Both front and back sides were patterned by conventional photolithography. SiN films (or AlN films) were patterned and etched with fluorine-based (or chlorine-based) inductively coupled plasma reactive-ion etching (ICP-RIE) to define the membrane geometry. The back-side $\rm{SiO_{2}}$ was then patterned and etched with fluorocarbon-based ICP-RIE. The Si wafer was etched from the back-side using patterned $\rm{SiO_{2}}$ as a mask in a deep reactive ion etch (DRIE) Bosch process, leaving approximately 50\,$\mu$m of Si below the patterned membrane region. Final membrane release was accomplished by a KOH (potassium hydroxide) wet etch for SiN membranes, and $\rm{XeF_2}$ (xenon difluoride) dry release for AlN membranes. Depending on the shape of a membrane hole, a specific shape of the membrane MOT can be prepared to interface the atoms through integrated photonics/electronics.

\section*{Trapping and sub-Doppler cooling of atoms in a sub-millimeter membrane hole}

\begin{figure}
\includegraphics[width=1\columnwidth]{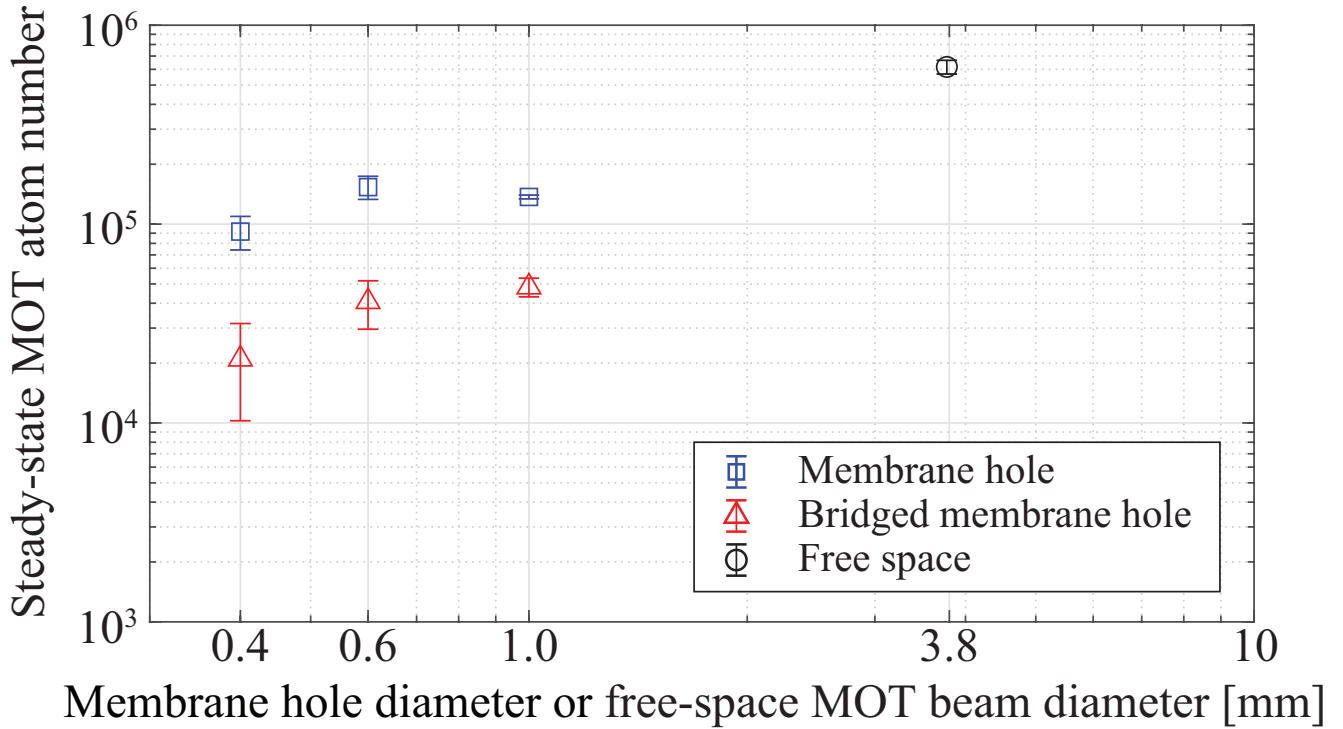}
\caption{Steady-state MOT atom number versus membrane hole diameter ($d_{hole}$) or free-space MOT beam diameter ($d_{MOT}$). The $d_{MOT}$ comes from the $1/e^2$ Gaussian beam diameter and the  $d_{hole}$ comes from the etched hole diameter of the membrane MOT devices. The six-beam MOTs were generated by a cooling beam and a repump beam in Cs atoms (852nm, D2 transition).  Steady-state MOT atom numbers were measured in the free space (black circle), the membrane hole (blue square), and the bridged membrane hole (red triangle) using an absorption probe (895nm, D1 transition). We used a  3.8\,mm beam diameter ($d_{MOT}$) for the free-space MOT and 0.4, 0.6, 1.0\,mm diameter membrane holes ($d_{Hole}$) for the membrane MOT and the bridged membrane MOT. The bridged membrane MOT experimentally simulates the situation of a 3-$\mu$m-width membrane ridge waveguide across the membrane hole.}
\label{fig_3}
\end{figure}

We characterized three different MOT configurations as follows: (1) free space, (2) unbridged membrane hole (see Fig.\,\ref{fig_1}(d): Left), and (3) bridged membrane hole (see Fig.\,\ref{fig_1}(d): Middle, Right)). Membrane MOT devices can trap atoms in sub-millimeter diameter holes and we compared steady-state atom number, loading rate, lifetime and sub-Doppler cooled atom temperature to the free-space MOT. The same laser cooling beams (852\,nm, Cs D2 transition) were used for all the MOT measurements with and without the membrane geometry affecting those characterization results. The localization of membrane MOT atoms is determined by the geometry of the membrane hole and the transparent membrane. The  transparent membrane ($5\,\rm{mm} \times 5\,\rm{mm}$) effectively provides a larger loading zone (i.e., MOT beams) compared to the membrane hole diameter but restricts atom trajectories. Therefore, most of the membrane MOT atoms are accumulated at the membrane hole ($d_{Hole} \leq 1 \,\rm{mm}$).

\begin{figure}
\includegraphics[width=1\columnwidth]{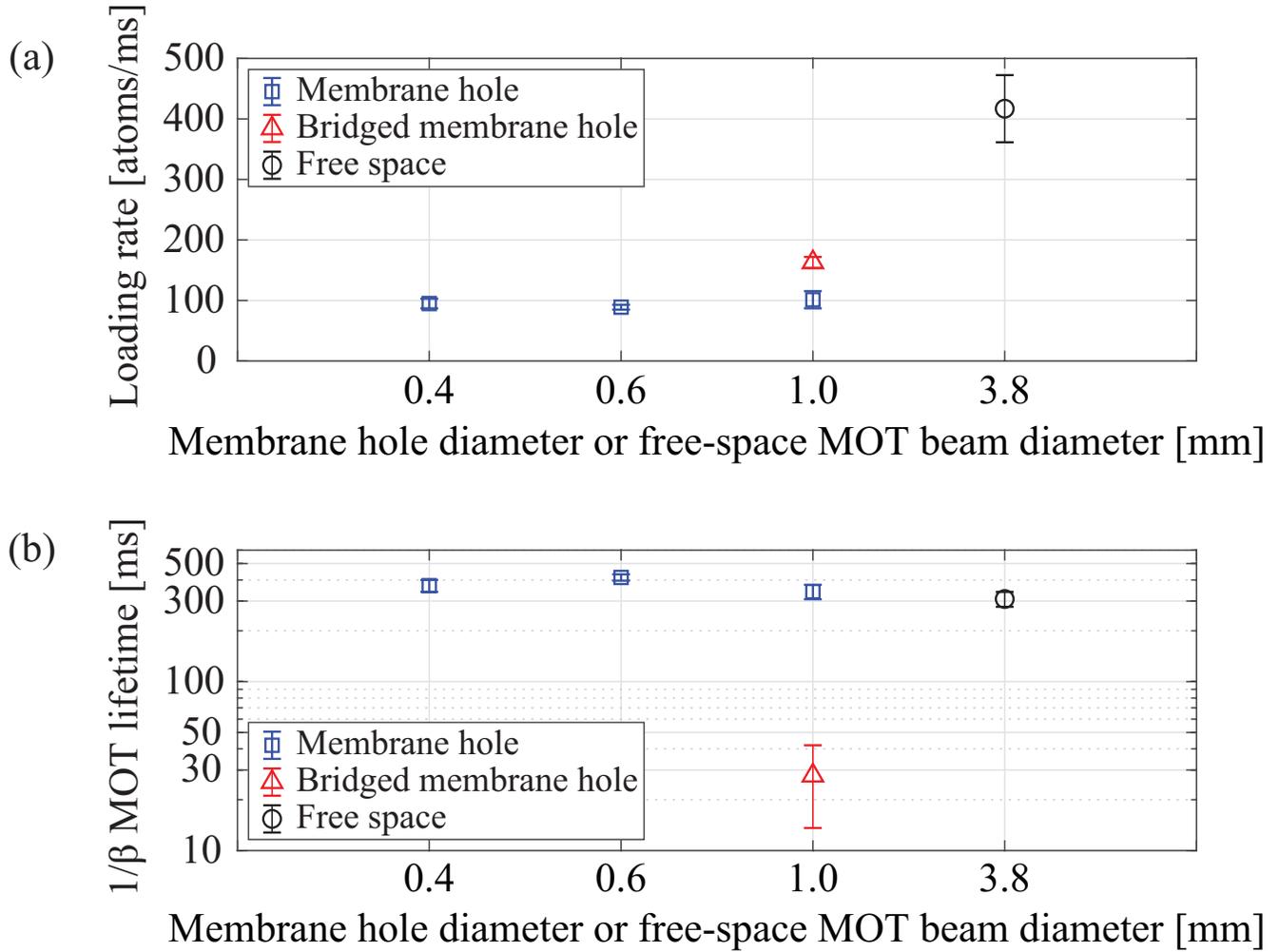}
\caption{(a) Loading rate versus membrane hole diameter or free-space MOT beam diameter. (b) MOT lifetime versus membrane hole diameter (or free-space MOT beam diameter). We measured the loading rates and lifetimes of generated MOT atoms in the free space (black circle), the membrane hole (blue square), and the bridged membrane hole (red triangle). Loading rate and MOT lifetime were measured with an absorption probe. We used 3.8\,mm diameter beams ($d_{MOT}$) for the free-space MOT (black circle) and 0.4, 0.6, 1.0\,mm-diameter membrane holes ($d_{Hole}$) for the membrane MOT (blue square) and the bridged membrane MOT (red triangle). The bridged membrane MOT experimentally simulates the situation of a 3-$\mu$m-width membrane ridge waveguide across the membrane hole.}
\label{fig_4}
\end{figure}

The measured steady-state atom number is shown in Fig.\,\ref{fig_3}. The MOT atom number at the membrane hole decreased by about 5-to-10 times compared to the free-space MOT ($d_{MOT} = 3.8\,\rm{mm}$), which may result from lower atom loading rate near the membrane device (see Fig.\,\ref{fig_4}(a)). Bridged membrane MOT devices with a comparable loading rate to the unbridged membrane devices show lower atom numbers than that of the membrane MOT devices at the membrane hole due to a lower $1/\beta$ MOT lifetime (see Fig.\,\ref{fig_4}(b)). The atom number was not strongly affected by the type of membrane (AlN or SiN). The transmittance of membranes should be maximized for a target wavelength, e.g., a laser cooling beam. Using 200nm-thick AlN and SiN membranes, the transmittance at 852nm wavelengths is greater than $95\%$, effectively providing two complete hemispherical capture volumes for Cs atoms. The transmittance depends on the thickness and refractive index of the membrane and the light wavelength. In addition, the membrane thickness and the height of the membrane ridge can affect the evanescent-field mode of the suspended membrane ridge waveguides\,\cite{Ayi-Yovo20}.

The optical intensity of Gaussian beams is defined as $I(r,z) = 2P/ (\pi w(z)^2) \exp(-2 r^2 / w(z)^2)$, and the beam radius $w(z)$ is the distance from the maximum intensity where the intensity drops to $1/e^2 (\approx 13.5\%)$. The optical power of each MOT beam is $P = 2.6 \pm 0.2\,\rm{mW}$, and the beam diameter of the MOT beam is $d_{MOT} = 2 w(z) \approx 3.8\,\rm{mm}$ in free space. Therefore, the free-space intensity of the MOT beam is $I = (41.7 \pm 3.2) \,I_{sat}$, where $I_{sat} = 1.1\,\rm{mW/cm^2}$ and $\Gamma = 2\pi \cdot 5.2\,\rm{MHz}$ (FWHM) for the $^{133}$Cs D2 transition at 852\,nm. The detuning of MOT beams is $\delta = - 2 \Gamma$ and the magnetic field gradient of the MOT coils is $dB/dz \approx 13.6\,\rm{G/cm}$. Using an absorption probe (895\,nm, Cs D1 transition), we measure $6.2 \times 10^5$ atoms in the free-space MOT ($d_{MOT} = 3.8\,\rm{mm}$), $9.2 \times 10^4$ atoms in the unbridged membrane MOT ($d_{Hole} = 0.4\,\rm{mm}$), and $2.1 \times 10^4$ atoms in the bridged membrane MOT ($d_{Hole} = 0.4\,\rm{mm}$; a 3-$\mu\rm{m}$-width dummy waveguide) as shown in Fig.\,\ref{fig_3}. 

The main difference between a free-space MOT and a membrane MOT is the localization of cold atoms. Most of the atoms in the membrane MOT exist within the boundary of a sub-millimeter hole due to the restriction of atom trajectories by the transparent membrane, which leads to the smaller size of a MOT cloud; the $1/e^2$-diameter of a MOT cloud is $180 \pm 6$\,$\mu$m for $d_{Hole} = 0.4\,\rm{mm}$. In the case of a free-space MOT, the localization of cold atoms needs to define a finite MOT capture volume with the MOT beam diameter. It is well known that the number of atoms accumulated in a MOT scales unfavorably with MOT beam diameter, especially below 2 mm where the scaling converts from $1/d^{3.6}$\,\cite{Chu92, Wieman92} to $1/d^6$\,\,\cite{Hoth13}. Hence, attempting to make a free-space MOT with a sub-millimeter $1/e^2$-diameter Gaussian beam results in a negligible number of atoms. The efficacy of the membrane MOT approach can be shown in a higher atom number localized within a membrane hole compared to the atom number in the free-space MOT specified by a cooling beam diameter which can localize atoms within the same size of the membrane hole. The membrane MOT is better at trapping atoms in a small volume compared to shrinking down the free-space MOT beams.

Based on the loading rate and lifetime measurement of the membrane and free-space MOTs, we found that the vacuum limited lifetime is the same near and far from the membrane in the case of no waveguide bridge, shown in Fig.\,\ref{fig_4}. The loading rate drops by about four near the membrane hole leading to the smaller atom number.  The loading rate is affected by the capture velocity, $v_c$, of the MOT\,\cite{Chu92}, and $v_c$ is a function of intensity, optical detuning, and diameter of the MOT beams, magnetic-field gradient, and the physical structure at the MOT. It is fair to assume that the membrane structure further limits $v_c$ and therefore reduce the loading rate as shown in Fig.\,\ref{fig_4}(a). In addition, the Cs vapor density in our experimental vacuum chamber is much lower than the saturation vapor pressure at room temperature. Under this condition, Cs vapor continuously deposit onto the membrane surface, which causes low near surface vapor density\,\cite{Harper09}. The MOT lifetime limited by collisions with background Cs atoms is therefore a bit longer, as shown in Fig.\,\ref{fig_4}(b). The bridged membrane hole decreases the MOT lifetime (see Fig.\,\ref{fig_4}(b)) due to the collision between cold atoms and the dummy waveguide, leading to a smaller steady-state atom number.

The usual MOT loading equation\,\cite{Foot92} is  $N(t) = \frac{\alpha}{\beta}(1-e^{-\beta t})$, where $\alpha$ is loading rate (atoms/ms) and $1/\beta$ is MOT lifetime (ms); $\beta$ is loss rate. The steady-state atom number of the membrane MOT, $\alpha/\beta$, is determined by loading rate and loss rate. Similar to a reference\,\cite{Hinds11}, the atom number in the MOT is significantly affected by the position of MOT atoms relative to the surface. Additional atom loss increases when the MOT approaches the surface because of atomic collisions with the surface. This attenuates the MOT atom number when the cloud comes within 300\,$\mu$m of the surface. For this case, the loading rate ($\alpha$) is constant (it is likely that the loading also decreases when the atoms are near a surface), whereas the loss rate ($\beta$) exhibits a dramatic increase near the surface.
 
The membrane MOT is always centered inside the hole. The distance to the cloud from the membrane hole edge is a half of the hole diameter such as 200, 300, 500\,$\mu$m for $d_{Hole} = 0.4, 0.6, 1.0\,\rm{mm}$ (photolithography mask pattern). As shown in Fig.\,\ref{fig_4}(a), loading rates of MOT atoms decrease by a factor of four near the membrane compared to the free-space MOT, which leads to the smaller steady-state atom number. In addition, the MOT lifetime of the bridged membrane MOT device (red triangle) is significantly lower than other cases, but other membrane MOT devices (blue square) have similar MOT lifetimes to the free-space MOT ($d_{MOT} = 3.8\,\rm{mm}$) MOT as shown in Fig.\,\ref{fig_4}(b). Collisions of atoms with the waveguide surface is a likely contributing factor in the higher loss rate. Other possibilities include a reduced effective capture volume (partially blocked MOT beams; reduced intensity and impure polarization of cooling beams, surface scattered photons with different k-vectors) and additional waveguide scattered photons with different k-vectors.
 
\begin{figure}
\includegraphics[width=1\columnwidth]{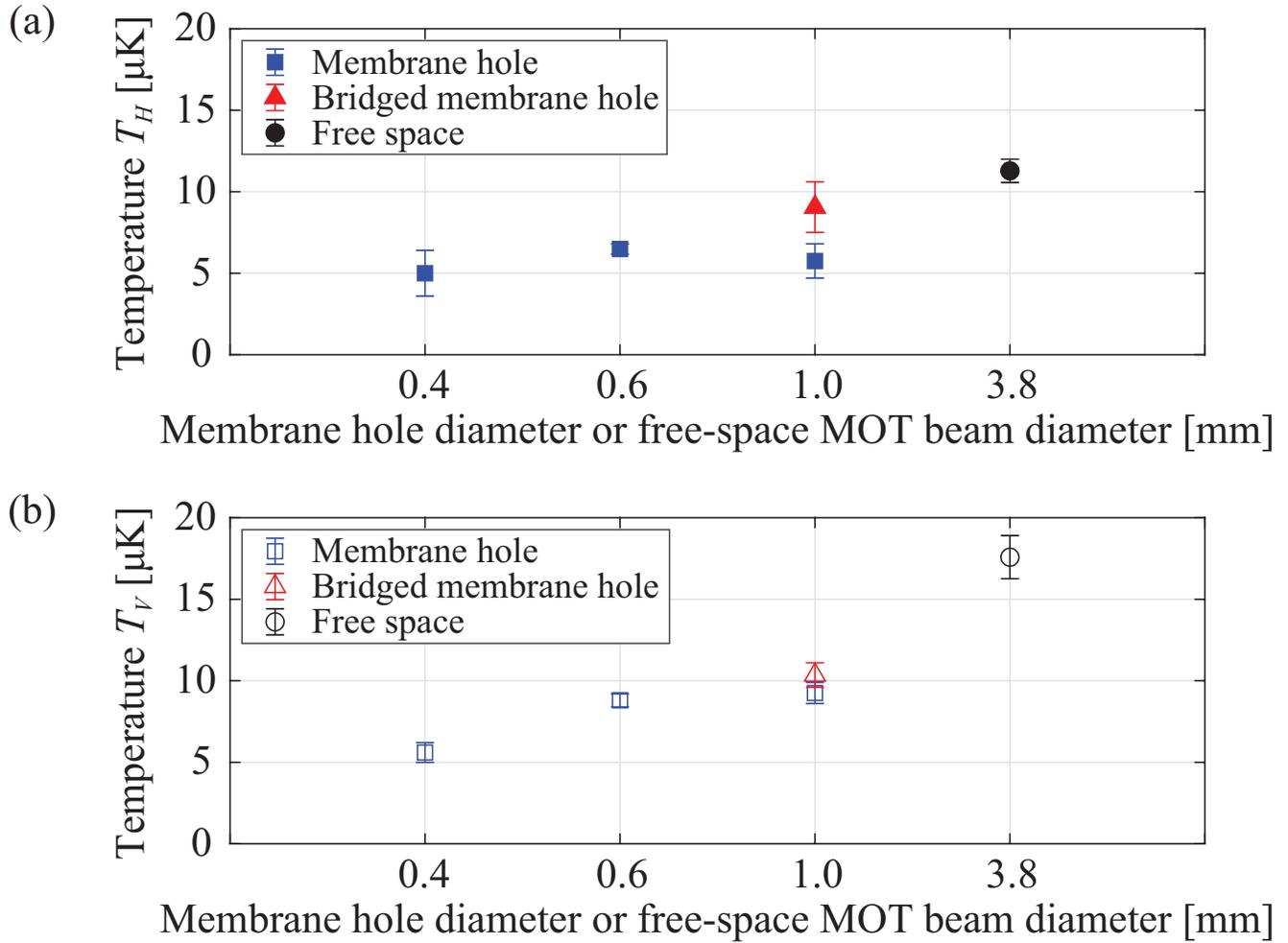}
\caption{Sub-Doppler cooled temperature of MOT atoms versus membrane hole diameter or free-space MOT beam diameter. (a) Temperature $T_H$, the temperature of atoms on the horizontal plane. (b) Temperature $T_V$, the temperature of atoms along the vertical axis. For this, we performed a millisecond of polarization gradient cooling. We used 3.8\,mm diameter beams ($d_{MOT}$) for the free-space MOT (black circle) and 0.4, 0.6, 1.0\,mm-diameter membrane holes ($d_{Hole}$) for the membrane MOT (blue square) and the bridged membrane MOT (red triangle). The bridged membrane MOT experimentally simulates the situation of a 3-$\mu$m-width membrane ridge waveguide across the membrane hole.}
\label{fig_5}
\end{figure}

We obtained sub-Doppler cooled temperatures of membrane MOT atoms as shown in Fig.\,\ref{fig_5}, which means the membrane MOT will be practical for real-world applications. Starting from steady-state MOTs, polarization gradient (PG) cooling sequence ($\sim 1 - 2\,\rm{ms}$) is performed with intensity lowering and frequency ramping while keeping the quadrupole magnetic field on. We measured atom temperature with time-of-flight measurement after sub-Doppler cooling. The temperatures of the free-space MOT atoms ($d_{MOT} = 3.8\,\rm{mm}$), membrane MOT atoms ($d_{Hole} = 0.4, 0.6, 1.0\,\rm{mm}$), and bridged membrane MOT atoms ($d_{Hole} = 1.0\,\rm{mm}$) are similar. The bridged section of the membrane MOT device imitates a 3-$\mu$m-width membrane ridge waveguide across the membrane hole. The horizontal temperature ($T_H$) corresponds to the temperature of atoms on the plane of four horizontal laser cooling beams, and the vertical temperature ($T_V$) corresponds to the temperature of atoms along the gravity axis of two vertical laser cooling beams as shown in Fig. 1 (b). Both horizontal and vertical temperature follow the trend of the time of flight (TOF) measurements even though the membrane MOT devices affect the expansion of atoms and there is some uncertainty in estimating the temperature inside the membrane. The temperature of membrane MOTs (blue square) shows a bit lower temperature than the free-space MOT (black circle). This could be due to hotter atoms colliding with the membrane and disappearing before the time-of-flight detection. The total atom number decreases, but we can achieve a colder cloud inside the hole. The lowest cloud temperature of the membrane MOT devices ($d_{Hole} = 0.4\,\rm{mm}$, blue square) is 5.3\,$\mu$K. The lowest measurable cloud temperature of the bridged membrane MOT devices ($d_{Hole} = 1.0\,\rm{mm}$, red triangle) is 9.7\,$\mu$K. In the temperature measurement of the sub-Doppler cooled membrane-MOT atoms, the atom number at the membrane hole appears to drop off with drop time as the MOT atoms expand and approach to the membrane. However, the $1/\beta$ MOT lifetime measurement of the Doppler cooled membrane-MOT atoms (Fig.\,\ref{fig_4}(b)) would not limit the atom number on the time scale (1-to-5\,ms) of the time-of-flight, temperature measurement. In particular, the MOT lifetime of the bridged membrane was lower than that of other membranes. This will require further study in future investigations.

\section*{Conclusion}
We developed membrane MOT devices capable of capturing $10^5$ cold atoms in a sub-millimeter diameter center hole of a transparent membrane. Sub-Doppler cooling in the membrane MOT was demonstrated with the temperature of 10\,$\mu$K. This membrane device can accumulate many atoms at the center of the membrane hole during the laser cooling process, and atoms can dissipate kinetic energy and relax into the center hole without entering the other hemisphere atom loading zone. This device was designed to improve the localization of cold atoms onto the suspended membrane ridge waveguide. By implementing the membrane hole, we achieve efficient atom loading around the suspended waveguide by leveraging two large hemispherical MOT capture volumes. This device can enable photonic atom trap integrated platforms (ATIP)\,\cite{Gehl21} with neutral atoms providing scalability, homogeneous physical properties, long coherence and lifetimes, and room-temperature operability. Membrane MOT devices with integrated photonics can utilize the guided, evanescent-field modes to trap and interface atoms. In addition, integrated photonics/electronics can be fabricated on the membrane device to enable advanced integration required for quantum applications.

\section*{Acknowledgement}
 We thank Peter Schwindt, Matt Eichenfield and Aleem Siddiqui for their support and helpful discussion. This work was supported by the Laboratory Directed Research and Development program at Sandia National Laboratories. Sandia National Laboratories is a multimission laboratory managed and operated by National Technology and Engineering Solutions of Sandia, LLC., a wholly owned subsidiary of Honeywell International, Inc., for the U.S. Department of Energy's National Nuclear Security Administration under contract DE-NA-0003525. This paper describes objective technical results and analysis. Any subjective views or opinions that might be expressed in the paper do not necessarily represent the views of the U.S. Department of Energy or the United States Government.

\section*{Author contributions statement}
Y.-Y. J., the principal investigator, conceived experiments and conducted initial experiments and simulations. J. L. and G. B. conducted the experiments and analyzed the results, and J.L. wrote the manuscript. J.M. and E.D. fabricated membrane MOT devices. All authors reviewed the manuscript.

\section*{Additional information}
\textbf{Competing interests}  The authors declare that they have no competing interests.

\end{document}